\documentstyle[aps,prl,multicol,epsf,psfig]{revtex}
\draft
\tightenlines

\begin{document}

\title{ A Measure of data-collapse for scaling}

\author{Somendra M. Bhattacharjee\cite{eml2}}
\address{Institute of Physics, Bhubaneswar 751 005, India\\ 
Dipartimento di Fisica, Universit\`a di Padova, Via
  Marzolo 8, 35131 Padova, Italy}
\author{Flavio Seno\cite{eml1}}
\address{INFM and Dipartimento di Fisica, Universit\`a di Padova, Via
  Marzolo 8, 35131 Padova, Italy}

\date{\today}
\maketitle

\widetext
\begin{abstract}
  Data-collapse is a way of establishing scaling and extracting
  associated exponents in problems showing self-similar or self-affine
  characteristics as e.g. in equilibrium or non-equilibrium phase
  transitions, in critical phases, in dynamics of complex systems and
  many others.  We propose a measure to quantify the nature of data
  collapse. Via a minimization of this measure, the exponents and
  their error-bars can be obtained.  The procedure is illustrated by
  considering finite-size-scaling near phase transitions and quite
  strikingly recovering the exact exponents.
\end{abstract}
%\pacs{05.10.-a,05.70.Jk, 03.65.-w,64.60.-i}

\begin{multicols}{2}

Scaling, especially finite size scaling (FSS), has emerged as an
important framework for understanding and analyzing problems involving
diverging length scales. Such problems abound in condensed matter,
high energy and nuclear physics, equilibrium and non-equilibrium
situations, thermal and non-thermal problems, and many more.  The
operational definition of scaling is this: A quantity $m(t,L)$
depending on two variables, $t$ and $L$, is considered to have scaling
if it can be expressed as
\begin{equation}
\label{eq:mt}
m(t,L)=L^d f(t/L^c).
\end{equation}
Depending on the nature of the problem of interest, $m$ may refer to
magnetization, specific heat, size or some other characteristic of a
polymer, width of a growing or fluctuating surface etc.  Eq.
\ref{eq:mt} is the FSS form if $L$ is a linear dimension of the system
and $t$ is any other variable, could even be time in dynamics. In the
thermodynamic limit of infinite-sized systems, such a scaling would
have $t$ and $L$ representing two thermodynamic parameters like
magnetic field, pressure, chemical potential etc or one could be time.
If $L$ is a length scale, then $d$ would look like the dimension of
this quantity $m$, and $c$ of variable $t$.  In fluctuation-dominated
cases, it is generally a rule, rather than an exception, that $d$ and
$c$ assume nontrivial values, different from what one expects from a
dimensional analysis.  The exponents and the scaling function $f(x)$
then characterize the behavior of the system. The fact that two
completely independent variables (both conceptually and as controlled
in experiments) combine in a nontrivial way to form a single one leads
to an enormous simplification in the description of the phenomenon.
This underlies the importance of scaling.

A quantitative way of showing scaling is data-collapse (also called
scaling plot) that goes back to the original observation of Rushbrooke
that the coexistence curves for many simple systems could be made to
fall on a single curve\cite{stanley}.  For example, the values of
$m(t,L)$ (Eq. \ref{eq:mt}) for various $t$ and $L$ can be made to
collapse on a single curve if $m L^{-d}$ is plotted against $t
L^{-c}$.  The method of data-collapse therefore comes as a powerful
means of establishing scaling.  It is in fact now used extensively to
analyze and extract exponents especially from numerical simulations.
Given the importance of scaling in wide varieties of problems, it is
imperative to have an appropriate measure to determine the ``goodness
of collapse'' - not to be left to the eyes of the beholder.

In this paper, we propose {\it a measure} that can be used to quantify
``collapse''.  This measure can be used, via a minimization principle,
for an automatic search for the exponents thereby removing the
subjectiveness of the approach.  To show the power of the method and
the measure, we use it for two exactly known cases, namely the
finite-size-scaling of the specific heat for (1) the one-dimensional
ferro-electric six vertex model\cite{lieb} showing a first order
transition \cite{smb_mri}, and (2) the Kasteleyn dimer
model\cite{kast} exhibiting the continuous anisotropic
Pokrovsky-Talapov transition\cite{pok,dimer}.  In addition, to show
the usefulness of the method in case of noisy data as expected in any
numerical simulation, we consider the one-dimensional case with extra
Gaussian noise added (by hand).  It is worth emphasizing that the
proposed procedure, without any bias, recovered the exactly known
exponents from the specific heat data for finite systems.

If the scaling function $f(x)$ of Eq. \ref{eq:mt} is known, then 
the sum of residuals
\begin{equation}
\label{eq:res1}
R = \frac{1}{N} \ \sum \mid L^{-d}\  m -f(t/L^c)\mid,
\end{equation}
where the sum is over all the data points, is minimum for the right choice
of $(d,c)$. In absence of any statistical or systematic error, the
minimum value is zero.

However in most situations the function itself is not known but is
generally an analytic function.  In case of a perfect collapse, any
one of the sets (say set $p$) can be used for $f(x)$.  An
interpolation scheme can then be used to estimate the values for other
sets in the {\it overlapping regions }. The residuals are then
calculated.  Since this can be done for any set as the basis, we
repeat the procedure for all sets. Let the tabulated values of $m$ and
$t$ be denoted by $m_{ij}, t_{ij}$ ( $i$th value of $t$ for the $j$th
set of $L$ (i.e.$L=L_j$ for set $j$ )). We now define a quantity $P_b$,
\begin{equation}
\label{eq:pb}
P_b=\left[\frac{1}{{\cal N}_{\rm over}}\ \sum_{p}\ \sum_{j\neq p}
\ \sum_{i,{\rm over}} \mid L_j^{-d}\  m_{i,j} - 
{\cal E}_p(L_j^{-c}\, t_{ij}) \mid^{q}\right]^{^{1/q}},
\end{equation}
where ${\cal E}_p(x)$ is the interpolating function based on the
values of set $p$ bracketing the argument in question (of set $j$).
The innermost sum over $i$ is done only for overlapping points
(denoted by the ``$i$, over''), ${\cal N}_{\rm over}$ being the number
of pairs. Though defined with a general $q$, we use $q=1$.  For 
${\cal E}_p(x)$, a 4-point polynomial interpolation can be used and if any
complex singularity is suspected a rational approximation may be used.
Extrapolations are avoided.  The minimum of this
function $P_b$ is zero\cite{comm} and is achievable in the ideal case
of perfect collapse with correct values of $(d,c)$, i.e.,
\begin{equation}
\label{eq:min}
P_b \geq P_b|_{\rm abs\  min} =0
\end{equation} 
This inequality can then be exploited and a minimization of $P_b$
over $(d,c)$ can be used to extract the optimal values of the
parameters.

In addition to the values of the exponents, estimates of errors can be
obtained from the width of the minimum. A simple approach is 
to take the quadratic part in the individual directions along the (d,c) plane.
From an expansion of $\ln P_b$ around the minimum at $(d_0,c_0)$, the width 
is estimated as
\begin{mathletters}
\begin{equation}
\label{eq:width}
\Delta d= \eta d_0 \left[2 \ln \frac{P_b(d_0\pm \eta
    d_0,c_0)}{P_b(d_0,c_0)}\right]^{^{-1/2}} ,
\end{equation}
 and
\begin{equation}
\label{eq:width2}
\Delta c= \eta c_0 \left[2 \ln \frac{P_b(d_0,c_0\pm \eta
    c_0)}{P_b(d_0,c_0)}\right ]^{^{-1/2}},
\end{equation}
\end{mathletters}
for a given $\eta$.  Choosing $\eta=1\%$, the final estimate for the
exponents would be $d_0\pm \Delta d,c_0\pm \Delta c$ with the error
bar reflecting the width of the minimum at $1\%$ level.

We now use the proposed method for different test cases. In order to
implement the program\cite{program}, we have used the routines of numerical
recipes\cite{recip}.  To calculate $P_b$, {\small POLINT} or {\small
  RATINT}  has been used for interpolation with {\small HUNT} to place
a point in the table.  For minimization, {\small AMOEBA} has been used
thrice to  locate the minimum, each time  using the current estimates
to generate a new triangle enclosing the minimum. In the examples
given below,  there was no need for more sophisticated
minimization routines, which could be needed  in case of subtle
crossover behaviors or with nearby minima.

Let us first consider the one-dimensional six-vertex model which shows
a first-order transition\cite{smb_mri}.  With the partition function
$Z=2+(2x)^N$ for $N$ sites with $x=\exp(-\epsilon/k_B T)$ as the
Boltzmann factor, $\epsilon$ being the energy of the high-energy
vertices, $T$ the temperature and $k_B$ the Boltzmann factor, the
specific heat can be computed exactly.  The first-order transition is
at $x=1/2$, for $N\rightarrow\infty$, with a $\delta$-function jump in
specific heat.  The $N$-dependent specific heat (per site),$c_N$, is given
by\cite{smb_mri}
\begin{equation}
\label{eq:vert1}
c_N= k_B (\ln x)^2 2 N \frac{(2x)^N}{[2+(2x)^N]^2},
\end{equation}
which for large $N$ and small $t=1-2x$ has the scaling form of Eq. 
\ref{eq:mt} with $d=1$, $c=-1$ and
\begin{equation}
\label{eq:vert2}
f(z)= k_B (\ln 2)^2 2  \frac{e^z}{(2+e^z)^2}.
\end{equation}
From the exact formula, Eq. \ref{eq:vert1}, data were generated for
$N=10, 30,50, 70$ and $90$, for various values of temperatures.  A
minimization of $P_b$ gave us the estimate $d=.997\pm
0.04,c=.98\pm.06$, with $P_b=0.56881E-01$.  The exponents are very
close to the exact ones. The error-bars or the width of the minimum is
to be interpreted as an indication of the presence of non-scaling
corrections.  To test this, we have generated data from the exact
scaling function of Eq. \ref{eq:vert2}. An unbiased minimization of
$P_b$ then gave $d=1\pm 0.004,c=-1\pm 0.004$ with 
$P_b=0.34876E-03$.  The smallness of the residue and of the errors (or
the width of the minimum) represents a good data collapse.
The nature of the data-collapse for both the cases  is shown
in Fig. 1.

\vbox{
\begin{figure}
\psfig{file=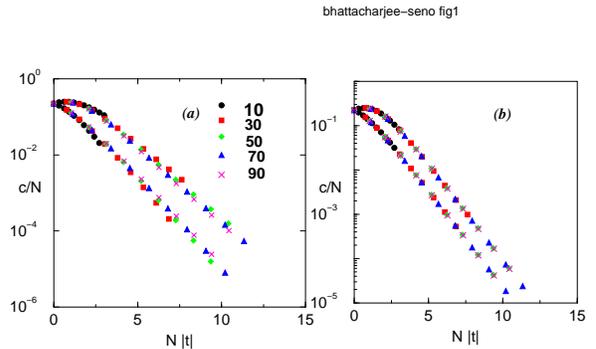,width=3.in} 
\narrowtext
\caption{The collapse of the specific heat for the 1-d vertex model as
  calculated from Eq. \ref{eq:vert1} [Fig.(a)] and from Eq.
  \ref{eq:vert2} [Fig. (b)].  The x-axis in both the plots is $\mid
  t\mid/N^c$. The upper (lower) branch is for $t>0$ ($t<0$).  For both,
  the exact values $d=1,c=-1$ are used.}
\label{fig:vert}
\end{figure}
}

A similar minimization of $P_b$ was carried out for the
two-dimensional Kasteleyn dimer model ( also isomorphic to 
a two-dimensional  5 vertex model).  This is an exactly solvable lattice 
model of the
continuous anisotropic Pokrovsky-Talapov transition for surfaces, and
shows a square-root singularity for specific heat with different
correlations lengths in the two directions\cite{dimer,comm2}.  The
specific heat for lattices of size $M$ along the direction of the
``walls'' and infinite in the transverse direction is known exactly
and its finite size scaling form has been discussed in Ref
\cite{dimer}.  Using the following formula for the specific heat per
site $c_M$
\begin{equation}
\label{eq:kast}
\frac{ c_M}{k_B a}= M \int_0^{2\pi} \ \frac{(2x \cos\phi)^M}{[1+(2x
\cos\phi)^M]^2}\ d\phi,
\end{equation}
specific heats data were generated for $M=10,30,50,70$ and $90$.  In
this formula, a few unimportant factors are put under $a$ and not
explicitly shown.  The critical point is at $x=1/2$. A  minimization of
$P_b$ gave $d=.5\pm 0.03,c=-.945\pm .02$ to be compared with the exact
values $d=.5,c=-1$.  The residue factor is $P_b=0.12424E-01$.  The
importance of correction terms are clear from Fig. 5 of Ref.
\cite{dimer}, and in our approach it gets reflected in the not too
small value of $P_b$.

The function $P_b$ for $q=1$ for the above two-dimensional problem is
shown as a surface plot over the 
$(d,c)$ plane in Fig. 1. The sharpness of the minimum is note-worthy.
In both the examples considered,  the performance of the method is
remarkable.

\vbox{
\begin{figure}
\psfig{file=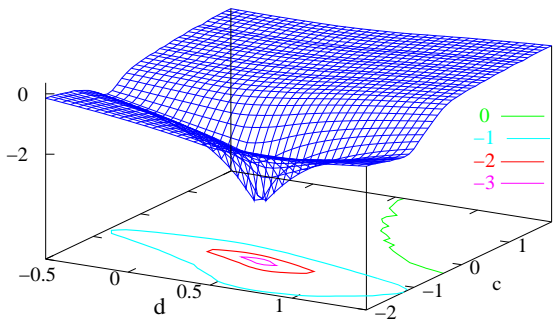,width=3in} 
\narrowtext
\caption{The residue $P_b$ with $q=1$ is shown over the $(d,c)$ plane. The 
  $z$-axis is in log scale.  A few contours
  of constant $\ln(P_b)$ are shown in different colors on the $(d,c)$ plane.}
\end{figure}
}

The last example we consider is the set of noisy data\cite{series}
where $c$ is calculated from Eq. \ref{eq:vert2} and Gaussian noise was
added to it so that $ c_{N,\eta}=| c_N\ (1+ A\eta)|$ where $\eta$
is a Gaussian deviate and $A$ is the amplitude of the noise added. The
absolute value is taken to keep $c_n$ positive.  The values of the
exponents are found to be insensitive to $A$ for $A<0.1$ and starts
changing for higher values of $A$.  In Fig. \ref{fig:noise} we show $
P_b$ against $A$.  The larger values of $P_b$ for larger $A$ is a sign
of poor collapse, as one finds by direct plotting with the estimated
values or exact values.

\vbox{
\begin{figure}
\psfig{file=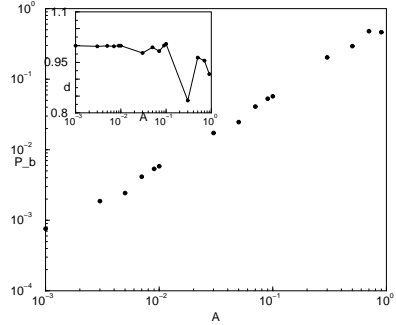,width=2in}
\narrowtext
\caption{Noisy data.  Plots of $P_b$  against $A$. Inset shows the
  estimated value of $d$ as a function of $A$.}
\label{fig:noise}
\end{figure}
}
To summarize, we have proposed a measure to quantify the nature of
data collapse in any scaling analysis of the form given by Eq.
(\ref{eq:mt}).  This measure can be used for an automated search for
the exponents.  The method is quite general and even-though we
formulated it in terms of power-laws as in Eq. 1, it can very easily
be adopted to other forms of scaling\cite{log}. We conclude that the
subjectiveness of data-collapse can be removed and $P_b$ could be used
as a quantitative measure to test or compare ``goodness of collapse''
in any scaling analysis.

We acknowledge financial support from MURST(COFIN-99).

 \end{multicols} 


\begin{references}
\bibitem{comm} Pathological cases of no overlap are avoided.
\bibitem{program} This program is available on
  request from the authors. It is flexible enough that the Numerical
  Recipes\cite{recip} routines could be replaced by any other suitable
  package.
\bibitem{comm2} Along the direction of the ``walls'' the bulk
  length-scale  exponent is  $\nu_y=1$ while in the transverse
  direction $\nu_x=1/2$.
\bibitem{log}. For example, when tested on a
log-type singularity as one finds for the specific heat in the
two-dimensional Ising model, the method does not give a good collapse
with power laws, but yields the correct results when the power-laws
are changed to log's.
\end{references}
\end{document}